\documentclass[noeprint,twocolumn,prb,showpacs,preprintnumbers,superscriptaddress,amsmath,amssymb,10pt,aps,floatfix]{revtex4-1}
\usepackage{graphicx}
\usepackage{dcolumn}
\usepackage{color}
\usepackage{bm}
\usepackage[varg]{txfonts}
\usepackage[hidelinks]{hyperref}
\hypersetup{
    colorlinks,
    citecolor=blue,
    filecolor=blue,
    linkcolor=blue,
    urlcolor=blue
}

\newcommand{\e}{\mathrm{e}}
\newcommand{\ud}{\,\mathrm{d}}
\newcommand{\im}{\mathrm{i}}
\newcommand{\br}{{\mathbf{r}}}
\newcommand{\vk}{{\vec{k}}}
\renewcommand{\Im}{\mathrm{Im}\,}
\DeclareMathOperator{\hh}{\mathcal{H}}
\DeclareMathOperator{\T}{\mathcal{T}}
\DeclareMathOperator{\F}{\mathcal{F}}

\begin{document}

\title{Dirac surface states in superconductors: a dual topological proximity effect}

\author{N. Sedlmayr}
\email{sedlmayr@umcs.pl}
\affiliation{Institute of Physics, M.~Curie-Sk{\l}odowska University, 20-031 Lublin, Poland}

\author{E. W. Goodwin} 
\affiliation{Department of Physics and Astronomy, Michigan State University, East Lansing, Michigan 48824, USA}

\author{M. Gottschalk} 
\affiliation{Department of Physics and Astronomy, Michigan State University, East Lansing, Michigan 48824, USA}

\author{I. M. Dayton}
\altaffiliation[Current address: ]{Northrop Grumman Corp., Baltimore, Maryland, USA}
\affiliation{Department of Physics and Astronomy, Michigan State University, East Lansing, Michigan 48824, USA}

\author{C. Zhang}
\affiliation{Department of Physics and Materials Research Laboratory, University of Illinois at Urbana-Champaign, Urbana, Illinois 61801, USA}

\author{E. Huemiller}
\affiliation{Department of Physics and Materials Research Laboratory, University of Illinois at Urbana-Champaign, Urbana, Illinois 61801, USA}

\author{R. Loloee}  
\affiliation{Department of Physics and Astronomy, Michigan State University, East Lansing, Michigan 48824, USA}

\author{T. C. Chasapis}
\affiliation{Department of Chemistry, Northwestern University, Evanston, Illinois 60208, USA}

\author{M. Salehi}
\affiliation{Department of Physics and Astronomy, Rutgers, The State University of New Jersey, Piscataway, New Jersey 08854, USA}

\author{N. Koirala}
\affiliation{Department of Physics and Astronomy, Rutgers, The State University of New Jersey, Piscataway, New Jersey 08854, USA}

\author{M. G. Kanatzidis}
\affiliation{Department of Chemistry, Northwestern University, Evanston, Illinois 60208, USA}
\affiliation{Materials Science Division, Argonne National Laboratory, Illinois 60439, USA}

\author{S. Oh}
\affiliation{Department of Physics and Astronomy, Rutgers, The State University of New Jersey, Piscataway, New Jersey 08854, USA}

\author{D. J. Van Harlingen}
\affiliation{Department of Physics and Materials Research Laboratory, University of Illinois at Urbana-Champaign, Urbana, Illinois 61801, USA}

\author{A. Levchenko}
\affiliation{Department of Physics, University of Wisconsin-Madison, Madison, Wisconsin 53706, USA}

\author{S. H. Tessmer} 
\email{tessmer@pa.msu.edu} 
\affiliation{Department of Physics and Astronomy, Michigan State University, East Lansing, Michigan 48824, USA}

\date{\today}

\begin{abstract}
In this paper we present scanning tunneling microscopy of  Bi$_2$Se$_3$ with superconducting Nb deposited on the surface. We find that the topologically protected surface states of the Bi$_2$Se$_3$ leak into the superconducting over-layer, suggesting a dual topological proximity effect. Coupling between theses states and the Nb states leads to an effective pairing mechanism for the surface states, leading to a modified model for a topological superconductor in these systems. This model is consistent with fits between the experimental data and the theory.
\end{abstract}

\maketitle

\section{Introduction}

The contact of materials with different long-range ordering modifies their properties near the interface. The superconducting proximity effect (PE), whereby superconducting correlations leak into neighboring materials, is a particular example of such general phenomena that has been comprehensively studied over the years (see Refs.~\onlinecite{Deutscher1969,Esteve1997,Buzdin2005,Beenakker2015} for reviews and references therein). It has been recently used as the basis for topological superconductors where Majorana bound states (MBS) are predicted to be found by placing a conventional superconductor (S) on top of a topological insulator (TI)~\cite{Fu2008}. This leads to the appropriate $p$-wave pairing, which is needed for a topological superconductor, for example Majorana bound states can nucleate at vortices in such hybrid proximity systems \cite{Ioselevich2012}. This scenario was successfully experimentally demonstrated to occur in a field-induced vortex of a topological insulator-superconductor Bi$_2$Te$_3$-NbSe$_2$ heterostructure \cite{Xu2015}. Here we report on an overlooked dual effect whereby the topologically protected surface states (TPSS) of the TI leak into the superconducting material when the two are in contact. This opens up new possibilities for observing MBS. This phenomenon has recently been reported in \onlinecite{Trang2020}, in which a Pb(111) thin film grown on TlBiSe$_2$ shows that the Dirac cone of the TlBiSe$_2$ substrate can be seen on the surface of the superconducting film in ARPES experiments. In contrast we use scanning tunneling microscopy to probe the samples, and compare to our phenomenological theory.

In this paper we investigate the effect of the TPSS in Bi$_2$Se$_3$ on the superconductor Nb which is placed on top. We demonstrate that, contrary to previous expectations, the TPSS are not confined to the boundary between the materials but spread also into the superconductor. Thus to find Majorana fermions in these hybrid systems there is a delicate range of thicknesses of the superconductor layer which are thick enough to be bulk superconductors, but not so thick as to destroy the two-dimensional (2D) nature and Dirac cone of the TPSS, such that the MBS will exist. Inside the superconducting over-layer there is a superconducting pairing effect for the TPSS mediated by their coupling to the native superconductor states. Thus inside the bulk superconductor one has an effective $p$-wave like topological superconductor, provided the superconducting layer is of the appropriate thickness. The combination of the TPSS and the standard Bardeen-Cooper-Schrieffer (BCS) superconductivity gives rise to a striking density of states profile which can be measured using a scanning tunneling microscope (STM). The induced superconductivity of the TPSS is nonetheless not equivalent to the proximity induced pairing which gives rise to the topological superconductor in Ref.~\onlinecite{Fu2008}.

\begin{figure}[t!]
\includegraphics[width=0.9\columnwidth]{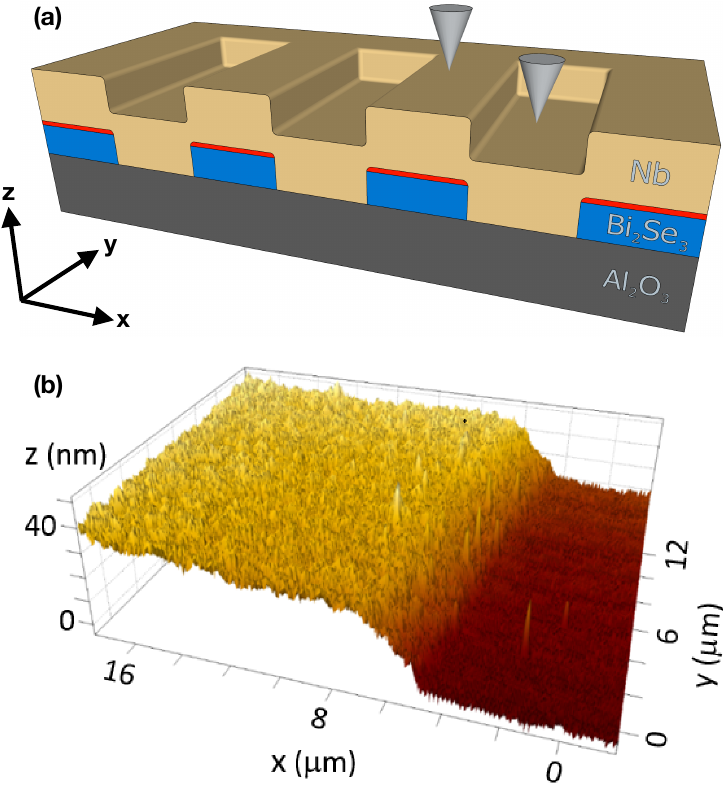} 
\caption{[Color online] (a) A schematic illustration for the geometry of the type A sample studied.  The alternating stripes of underlying Bi$_2$Se$_3$ allowed us to measure the superconductor Nb in contact with two different underlying layers: Al$_2$O$_3$ and Bi$_2$Se$_3$.  Density of states measurements with a Al$_2$O$_3$ sub-layer showed a typical superconducting energy gap [see figure \ref{ref:figure2}(c)].  The same measurement with a Bi$_2$Se$_3$ sub-layer showed indications of the dual topological proximity effect [see figure \ref{ref:figure2}(b)]. (b) Atomic force microscopy (AFM) topograph showing the boundary between alternating stripes.  The height difference between the Bi$_2$Se$_3$ sub-layer and Al$_2$O$_3$  sub-layer indicates a height of 40 nm for the underlying Bi$_2$Se$_3$, consistent with 40 quintuple layers.  The RMS roughness of the Bi$_2$Se$_3$ sub-layer is approximately 4 nm.}
\label{ref:figure1}
\end{figure}

After the prediction of $p$-wave pairing present in hybrid S-TI systems, there have been many other theory proposals and a wide range of experiments aiming to reveal the symmetry implications of this state in various observables e.g. Josephson current-phase relationship, tunneling conductance, current noise spectra. In particular, some experiments have focused on thin films of Bi$_2$Se$_3$ in proximity to an $s$-wave \cite{Wang2012b,Koren2012} or $d$-wave \cite{Wang2013a} superconductor, or on other TI thin films \cite{Xu2014}. Naturally such a system can only exhibit true 2D TPSS when it is thick enough to be approximately considered as a bulk 3D material, but signatures of $p_x+ip_y$ pairing induced by the superconducting proximity effect and Majorana bound states are thought to be present. Here we present results on a superconducting layer either on the surface of a bulk TI, Bi$_2$Se$_3$, or on the surface of an insulator, Al$_2$O$_3$; see figure \ref{ref:figure1} for an illustration of our devices. The primary experiment tests contact with the sub-layer TI (Bi$_2$Se$_3$); the Al$_2$O$_3$ sub-layer stripes allow for a control measurement done during the same data run, using the same tip, and all other identical testing conditions. The surface of the TI becomes superconducting due to the proximity effect, which may show signs of $p$-wave superconductivity \cite{Zareapour2012,Zhang2011b,Finck2014,Finck2016,Dayton2016,Banerjee2018,Alspaugh2018,Yang2019a}. It was found that a long range proximity effect is present for the TPSS \cite{Dayton2016}. Furthermore the density of states displays oscillations in space and energy which are reminiscent of, though strictly speaking distinct from, Tomasch and Friedel oscillations \cite{Wolfram1968}. Signatures of topological superconductivity and MBS have also been seen in point contact \cite{Sasaki2011} and transport experiments \cite{Sacepe2011,Veldhorst2012,Qu2012,Williams2012,Cho2013,Kurter2014,Kurter2014a,Sochnikov2015,Stehno2016}.

We present data from experiments on two samples, A and B. Samples A show the effect of the TPSS and of the superconducting gap separately, and allows us also to consider superconductors both with and without the underlying TI. Samples of type B have these salient features superimposed, and allow us to compare directly with theoretical predictions of a phenomenological model of the system.

\section{Samples and measurements}

The TI used in this experiment is Bi$_2$Se$_3$.  Five atomic planes with atomic order Se$^1$-Bi$^1$-Se$^2$-Bi$^1$-Se$^1$ form a quintuple layer (QL); the QLs are weakly bound to each other, making it possible to readily expose a pristine surface for study.  The exposed QL supports the existence of the TPSS, which features a single Dirac cone.

Two distinct sample growth methods were employed.  Samples of type A consist of 40 QL of Bi$_2$Se$_3$ grown via molecular beam epitaxy on c-plane sapphire~\cite{Bansal2012}.  Via mechanical masking and Ar+ ion milling, stripes of Bi$_2$Se$_3$ were removed, exposing bare Al$_2$O$_3$.  In another deposition step, the samples were gently milled before evaporating Nb of 40 or 60 nm on the entire sample.  A capping layer of 5 nm of Au was evaporated in-situ with the Nb layer to prevent the formation of NbOx.  This process was used both for the experimental stripes (Nb on Bi$_2$Se$_3$) and for the control stripes (Nb on Al$_2$O$_3$).
  
Samples of type B are uniform with respect to the x-y plane. They were grown by slowly cooling a stoichiometric mixture of Bi and Se from a temperature of $850^\circ$C.  The surface of the crystal was then cleaved in a Nitrogen gas environment. Recent work performed in our lab shows cleaving Bi$_2$Se$_3$ in a nitrogen gas environment shifts the Dirac cone towards the Fermi level~\cite{Gottschalk2020}; a similar effect has been seen with water vapor using ARPES and has been predicted for other gases~\cite{Benia2011}. Subsequently, 30 nm of Nb were dc sputtered on the surface at room temperature.  Samples were then transferred to our custom-designed Besocke-style STM system for measurement.  All the previous steps were done in a vacuum, nitrogen, or helium environment, so the sample is minimally exposed to air. To measure the tunneling spectra, the dc voltage applied between the tip and sample was summed with a 100 Hz sinusoidal voltage of 0.3 mV rms for most of the data shown in this paper. However, the  data shown in figure \ref{ref:figure2}(b) utilized an ac amplitude of 4.0 mV rms. The tip was positioned over the area of interest and the STM feedback turned off; the dc voltage was then slowly ramped allowing $dI/dV$-versus-$V$ to be measured using a standard current amplifier and lock-in amplifier. All STM spectra presented were taken at 4.2 K.

\section{Theoretical model}

To model our experimental system we consider a superconductor  deposited on top of a 3D topological insulator, as in figure \ref{ref:figure1}. The superconductor is a metal with $s$-wave pairing, but the surface states from the TI will also spread into the metal, a process which is inevitable based on generic considerations. We want to stress that this is the default behaviour of the states. In the absence of any energy barrier the TPSS states will leak into any adjacent material. This can be seen in numerical simulations, and similar effects are sen in topological superconducting wires.\cite{Guigou2016,Ptok2017} We seek for the simplest possible model that captures this physics. Therefore the minimal Hamiltonian has four terms  
\begin{equation}\label{ham1}
H=H_{\rm BCS}+H_{\rm TPSS}+H_{\rm C}\,,
\end{equation}
where $H_{\rm BCS}$ is the BCS Hamiltonian for a simple clean 2D metal including $s$-wave pairing, $H_{\rm TPSS}$ is for the TI surface states (TPSS), which have spread throughout the Nb over-layer, and $H_{\rm C}$ is a two-particle local coupling between the surface states and metallic states.

Firstly, the superconductor, in our case Nb, is described by a BCS Hamiltonian $H_{\rm BCS}=\int\ud^2\br\Psi^\dagger_\br\hh_{\rm BCS}\Psi_\br$ with $\br=(x,y)$ the 2D spatial coordinate, and 
\begin{equation}
\hh_{\rm BCS}=\hat{\xi}{\bm \tau}^z+\Delta{\bm \tau}^x\,.
\end{equation}
We use the Nambu basis, with $\Psi^\dagger_\br=\{c^\dagger_{\br\uparrow},c^\dagger_{\br\downarrow},c_{\br\downarrow},-c_{\br\uparrow}\}$, and a wavefunction $\psi^T_\br$: $\{u_{\br\uparrow},u_{\br\downarrow},v_{\br\downarrow},v_{\br\uparrow}\}$. Here $c^\dagger_{\br\sigma}$ creates a particle of spin $\sigma$ at position $\br$. We will also use $\vec{\bm\tau}$ as the Pauli matrices acting in the particle-hole subspace and $\vec{\bm\sigma}$ as the Pauli matrices operating in the spin subspace. The band operator is $\hat\xi=-1/(2m) \nabla^2-\mu$ in two dimensions (2D).

The TPSS are described by \cite{Zhang2009}
\begin{equation}
\hh_{\rm TPSS}=\left(-\im v_F\nabla\cdot\vec{\bm \sigma}-\mu_{\rm TPSS}\right){\bm \tau}^z\,.
\end{equation}
with $\chi^\dagger_\br=\{a^\dagger_{\br\uparrow},a^\dagger_{\br\downarrow},a_{\br\downarrow},-a_{\br\uparrow}\}$ and $H_{\rm TPSS}=\int\ud \br\chi^\dagger_\br\hh_{\rm TPSS}\chi_\br$, where $a^\dagger_{\br\sigma}$ creates a particle of spin $\sigma$ at position $\mathbf{r}$ on the TI surface. 

We consider the simplest uniform coupling mechanism
\begin{equation}
H_{\rm C}=\gamma\int\ud^2\br\chi^\dagger_\br{\bm \tau}^z\Psi_\br+\textrm{H.c.}\,,
\end{equation}
namely a local spin independent hybridization of strength $\gamma$.

A Fourier transform, and a spin rotation which diagonalizes $H_{\rm TPSS}$ but leaves $\hh_{\rm M}+\hh_\Delta$ unaffected, decouples this Hamiltonian into two independent parts.
Let $\widetilde\Psi_\vk=\mathcal{U}_k^\dagger\mathcal{V}_\vk^\dagger\Psi_\vk$ and $\widetilde\hh_{\rm BCS}=\mathcal{U}_k^\dagger\mathcal{V}_\vk^\dagger\hh_{\rm BCS}\mathcal{V}_\vk\mathcal{U}_k$ with $\mathcal{U}_k$ and $\mathcal{V}_\vk$ two rotation matrices. Firstly we have
\begin{equation}
 \mathcal{U}_k=\frac{1}{\sqrt{2}}
 \begin{pmatrix}
 0&\frac{\Delta}{\alpha^-_k}&0&-\frac{\Delta}{\alpha^+_k}\\
 \frac{\Delta}{\alpha^-_k}&0&-\frac{\Delta}{\alpha^+_k}&0\\
 0&\frac{\alpha^-_k}{\epsilon_k}&0&\frac{\alpha^+_k}{\epsilon_k}&\\
  \frac{\alpha^-_k}{\epsilon_k}&0&\frac{\alpha^+_k}{\epsilon_k}&0
 \end{pmatrix}\,,
\end{equation}
where $\epsilon_k=\sqrt{\Delta^2+\xi_k^2}$ and $\alpha^\pm_k=\sqrt{\epsilon_k^2\pm\epsilon_k\xi_k}$. Secondly $\mathcal{V}_\vk$ is a spin rotation which commutes with $\hh_{\rm M}+\hh_\Delta$. Then we find
\begin{equation}\label{bcs}
\widetilde\hh_{\rm BCS}=\epsilon_k{\bm \tau}^z\,.
\end{equation}
For the TPSS we make the spin rotation $\widetilde\chi_\vk=\mathcal{V}_\vk^\dagger\chi_\vk$ with
\begin{equation}
\mathcal{V}_\vk=\e^{\frac{\im\phi_k{\bm \sigma}^z}{2}}
\e^{\frac{\im\pi{\bm \sigma}^x}{4}}\,.
\end{equation}
$\phi_k=\pi/2-\tan^{-1}[k_x/k_y]$ is the polar angle.
The rotated Hamiltonian density for the TPSS is
\begin{equation}
 	\widetilde\hh_{\rm TPSS}=\mathcal{V}_\vk^\dagger\hh_{\rm TPSS}\mathcal{V}_\vk=(v_Fk{\bm\sigma}^z-\mu_{\rm TPSS}){\bm \tau}\,.^z
\end{equation}
The coupling becomes
\begin{equation}\label{coupling}
\widetilde H_{\rm C}=\int\ud^2\vk\widetilde\chi^\dagger_\vk\underbrace{\gamma{\bm \tau}^z\mathcal{U}_k}_{={\bm \Gamma}_k}\widetilde\Psi_\vk+\textrm{H.c.}\,.
\end{equation}
In this basis the Hamiltonian Eq.~\eqref{ham1} can then be directly decoupled into two Hamiltonian densities:
\begin{equation}\label{ham2}
\hh^\pm=\begin{pmatrix}
-\xi_k&\Delta&0&-\gamma\\
\Delta&\xi_k&\gamma&0\\
0&\gamma&\zeta_{\pm k}&0\\
-\gamma&0&0&-\zeta_{\pm k}
\end{pmatrix}\,,
\end{equation}
with  $\zeta_k=v_Fk-\mu_{\rm TPSS}$. Diagonal blocks describe S and TI states, whereas off-diagonal terms describe their mutual coupling. 

From Eq.~\eqref{ham2} it is simple to find the dispersion of the eight energy bands:
\begin{eqnarray}\label{spectrum}
&&\varepsilon^{abc}_k=\frac{a}{\sqrt{2}}
\left[
2\gamma^2+\epsilon_k^2+{c\zeta_{ck}}^2+\phantom{\frac{}{}}\right. \nonumber \\ 
&&\left. b\sqrt{\left[\epsilon_k^2-{c\zeta_{ck}}^2\right]^2+4\gamma^2\left[\Delta^2+\left(\xi_k+c\zeta_{ck}\right)^2\right]}\right]^{1/2}\,,
\end{eqnarray}
where $a,b,c$ are each $\pm1$ and $\epsilon_k=\sqrt{\Delta^2+\xi_k^2}$. In the limit $\gamma\to 0$ we recover the BCS and TPSS dispersions as required.
For large coupling $\gamma$ there is still a full gap in the spectrum. However for small $\gamma$ there are states inside the BCS gap caused by the TPSS which will have only a weak superconducting pairing effect. We use this model to compare directly with the results of the STM measurements.

The density of states, where the delta function peaks have been broadened into Gaussians of width $\Gamma$, is given by
\begin{equation}
\nu(\omega)=\int\frac{\ud^2k}{4\pi^2}
\frac{\e^{-\frac{(\omega-\varepsilon_{\vec{k}})^2}{\Gamma^2}}}{\sqrt{\pi}\Gamma}  \,.
\end{equation}
This can be calculated numerically with Eq.~\eqref{spectrum}. This density of states prediction will break down once either the metallic density of states or the TPSS spectrum are no longer symmetric around zero energy.

\subsection{Effective model of topologically protected surface states with s-wave pairing}\label{sec_dos}

If we are interested in the properties of the TPSS we can find an effective model by integrating out the BCS states. In the functional integral representation this amounts to calculating
\begin{equation}
\e^{-S'}=\left\langle \e^{-T\sum_n\int\ud^2\vk\left[\widetilde\chi^\dagger_{n\vk}{\bm \Gamma}_k\widetilde\Psi_{n\vk}+\textrm{H.c.}\right]}\right\rangle_{\rm BCS}\,.
\end{equation}
We use the Matsubara formalism, with frequencies $\omega_n=2\pi(n+\frac{1}{2})$ for $n\in\mathbb{Z}$, and temperature $T$.
As the action we need to integrate over is quadratic this is a standard integral and we find
\begin{equation}
S'=\frac{1}{2}T\sum_{n}\int\ud^2\vk
\left[\widetilde\chi^\dagger_{n\vk}{\bm \Gamma}_k^\dagger {\bm g}_{nk}{\bm \Gamma}_k\widetilde\chi_{n\vk}+\textrm{H.c.}\right]\,,
\end{equation}
where ${\bm g}_{nk}=\langle\widetilde\Psi_{nk}\widetilde\Psi^\dagger_{nk}\rangle$ is the Green's function for the superconducting states. This is a correction to $\hh_{\rm TPSS}$ and we have an effective Hamiltonian density
\begin{equation}
\hh^{\rm eff}_{\rm TPSS}=\hh_{\rm TPSS}+\Delta^{\rm eff}_{nk}{\bm \tau}^x-\mu'_{nk}{\bm \tau}^z\,,
\end{equation}
where
\begin{equation}
\Delta^{\rm eff}_{nk}=\frac{\gamma^2\Delta}{\omega_n^2+\epsilon_k^2}\,, \textrm{ and }
\mu'_{nk}=\frac{\gamma^2\xi_k}{\omega_n^2+\epsilon_k^2}\,.
\end{equation}
Thus in the static limit we are left with the effective model:
\begin{eqnarray}
\mathcal{H}^{\rm eff}&=&(v_Fk{\bm\sigma}^z-\mu^{\rm eff}_{k}){\bm \tau}^z+\Delta^{\rm eff}_{k}{\bm \tau}^x\,,\\\nonumber
\Delta^{\rm eff}_{k}&=&\frac{\gamma^2\Delta}{\Delta^2+\xi^2_k}\,, \textrm{ and}\quad
\mu^{\rm eff}_{k}=\mu_{\rm TPSS}+\frac{\gamma^2\xi_k}{\Delta^2+\xi_k^2}\,.
\end{eqnarray}
If we can neglect the momentum dependence of the effective pairing, $\Delta^{\rm eff}_{k}\to\Delta$, and chemical potential, $\mu^{\rm eff}_{k}\to\mu$, then one is left with the Fu-Kane model\cite{Fu2008},
\begin{equation}\label{hfk}
\mathcal{H}_{\rm FK}=(v_Fk{\bm\sigma}^z-\mu){\bm \tau}^z+\Delta{\bm \tau}^x\,.
\end{equation}
The density of states of this model can be calculated analytically.

We start from Gorkov's equation:
\begin{equation}\label{gti}
\begin{pmatrix}
\im\omega_n-H & {\bm \Delta}\\
-{\bm \Delta}^\dagger & -\im\omega_n-H^* 
\end{pmatrix}\begin{pmatrix}
G_{n,\vec k}\\
\F^\dagger_{n,\vec k}
\end{pmatrix}
=
\begin{pmatrix}
1\\
0
\end{pmatrix}\,,
\end{equation}
with
\begin{eqnarray}
H^{(*)}&=&v_{F}\begin{pmatrix}
0&\pm k_x-\im k_y\\
\pm k_x+\im k_y&0
\end{pmatrix}-\mu\mathbb{I}_2\,,
\end{eqnarray}
which describes the 2D surface states of a 3D topological insulator. $\mathbb{I}_n$ is an $n\times n$ identity matrix. ${\bm \Delta}=\Delta i{\bm \sigma}^y$ is the s-wave pairing. Note that $\Delta H=H^*\Delta$. Naturally this is in the Matsubara representation with Green's functions
\begin{eqnarray}\label{matg}
G(\br,\tau;\br',\tau')&=&-\langle\T_\tau\psi_{\sigma}(\br,\tau)\psi^\dagger_{\sigma'}(\br',\tau')\rangle\\\nonumber
&=&T\sum_n \e^{-\im\omega_n(\tau-\tau')}G_n(\br,\br')\,,
\end{eqnarray}
and
\begin{eqnarray}\label{matf}
\F^\dagger(\br,\tau;\br',\tau')&=&-\langle\T_\tau\psi^\dagger_{\sigma}(\br,\tau)\psi^\dagger_{\sigma'}(\br',\tau')\rangle\\\nonumber
&=&T\sum_n \e^{-\im \omega_n(\tau-\tau')}\F^\dagger_n(\br,\br')\,,
\end{eqnarray}
where $\T_\tau$ is time ordering along the imaginary time axis and $\psi_{\sigma}(\br,\tau)$ is a Heisenberg operator. Finally $\br=(x,y)$ is the 2D spatial coordinate.

We find as the differential equation for $\F^\dagger$
\begin{equation}
\frac{\im{\bm\sigma}^y}{\Delta}\left[|{\bm \Delta}|^2+\omega_n^2+(H^*)^2\right]\F^{\dagger}_{n,\vec k}=1\,.
\end{equation}
The bulk solution is therefore
\begin{equation}
\F^{\dagger}_{n,\vec k}=-\frac{\im\Delta{\bm \sigma}^y}{\epsilon_{nk}^4-4\mu^2 v_F^2k^2}\begin{pmatrix}
\epsilon_{nk}^2&-2\mu v_Fk_+\\
-2\mu v_Fk_-&\epsilon_{nk}^2
\end{pmatrix}\,,
\end{equation}
where $k_\pm=k_x\pm \im k_y$ and
\begin{equation}
\epsilon_{nk}^2=\mu^2+\omega_n^2+\Delta^2+v_F^2k^2\,.
\end{equation}
The normal Green's function is found from
\begin{equation}
G_{n,\vec k}=-(\im\omega_n+H^*)\frac{i{\bm\sigma}^y}{\Delta}\F^{\dagger}_{n,\vec k}\,,
\end{equation}
which leads to
\begin{equation}
G_{n,\vec k}=-\frac{\im\omega_n+H^*}{\epsilon_{nk}^4-4\mu^2 v_F^2k^2}\begin{pmatrix}
\epsilon_{nk}^2&-2\mu v_Fk_+\\
-2\mu v_Fk_-&\epsilon_{nk}^2
\end{pmatrix}\,.
\end{equation}

The density of states for the TPSS is
\begin{equation}
\nu_{\rm FK}(\omega)=-\frac{1}{\pi}\Im \int\frac{d^2\vec k}{4\pi^2}\,\mathrm{tr} \left.G_{n,\vec k}\right|_{\im\omega_n\to\omega+\im\delta}\,,
\end{equation}
and substituting in $G_{n,\vec k}$ we are left with the integral
\begin{equation}
\nu_{\rm FK}(\omega)=\frac{2\omega}{\pi}\Im \int_{0}^\infty\frac{d k}{2\pi}\left.\frac{\epsilon_{nk}^2k}{\epsilon_{nk}^4-4\mu^2 v_F^2k^2}\right|_{\im\omega_n\to\omega+\im\delta}\,.
\end{equation}
This integral can be performed using standard techniques and we find
\begin{equation}\label{dosfk}
\nu_{\rm FK}(\omega)=\frac{|\omega|}{\pi v_F^2}\Theta(\omega^2-\Delta^2-\mu^2)\left[1+\frac{|\mu|\Theta(\omega^2-\Delta^2)}{\sqrt{\omega^2-\Delta^2}}
\right]\,,
\end{equation}
where $\Theta$ is the Heaviside theta function. We note that neither this nor the BCS density of states are sufficient for explaining the experimental results, which requires the fully coupled model.

\subsection{Effective model of s-wave superconductor in presence of spin-orbit coupling}\label{sec_eff_bcs}

We can also consider the effect on the BCS superconductor caused by the TPSS by integrating out the TPSS. In this case the self energy term would be
\begin{equation}
\e^{-S''}=\left\langle \e^{-T\sum_n\int\ud^2\vk\left[\widetilde\chi^\dagger_{n\vk}{\bm \Gamma}_k\widetilde\Psi_{n\vk}+\textrm{H.c.}\right]}\right\rangle_{\rm TPSS}\,.
\end{equation}
After a standard calculation this leads to the effective Hamiltonian
\begin{equation}
\hh^{\rm eff}_{\rm BCS}=\hh_{\rm BCS}+\left[\Sigma^{\rm eff}_{nk}-\alpha^{\rm eff}_{nk}k{\bm\sigma}^y\right]{\bm\tau}^z\,.	
\end{equation}
where
\begin{eqnarray}
 \Sigma^{\rm eff}_{nk}&=&\frac{\gamma^2\mu(\mu^2+\omega_n^2-v_F^2k^2)}{4\mu^2\omega_n^2+[\mu^2-\omega_n^2-v_F^2k^2]^2}\,,\textrm{ and}\\\nonumber
 \alpha^{\rm eff}_{nk}&=&\frac{\gamma^2v_F(\mu^2-\omega_n^2-v_F^2k^2)}{4\mu^2\omega_n^2+[\mu^2-\omega_n^2-v_F^2k^2]^2}\,.
\end{eqnarray}
$\Sigma^{\rm eff}_{nk}$ and $\alpha^{\rm eff}_{nk}$ describe electronic and spin-orbit scattering events respectively, which will contribute to the broadening of the BCS density of states. Again, this description leads, assuming the variation with momentum is small and taking the static limit, to a well known model for a topological superconductor\cite{Lutchyn2010,Oreg2010}.

\section{Results and discussion}

\subsection{Samples of Type A}

\begin{figure}[t!]
\includegraphics[width=0.99\columnwidth]{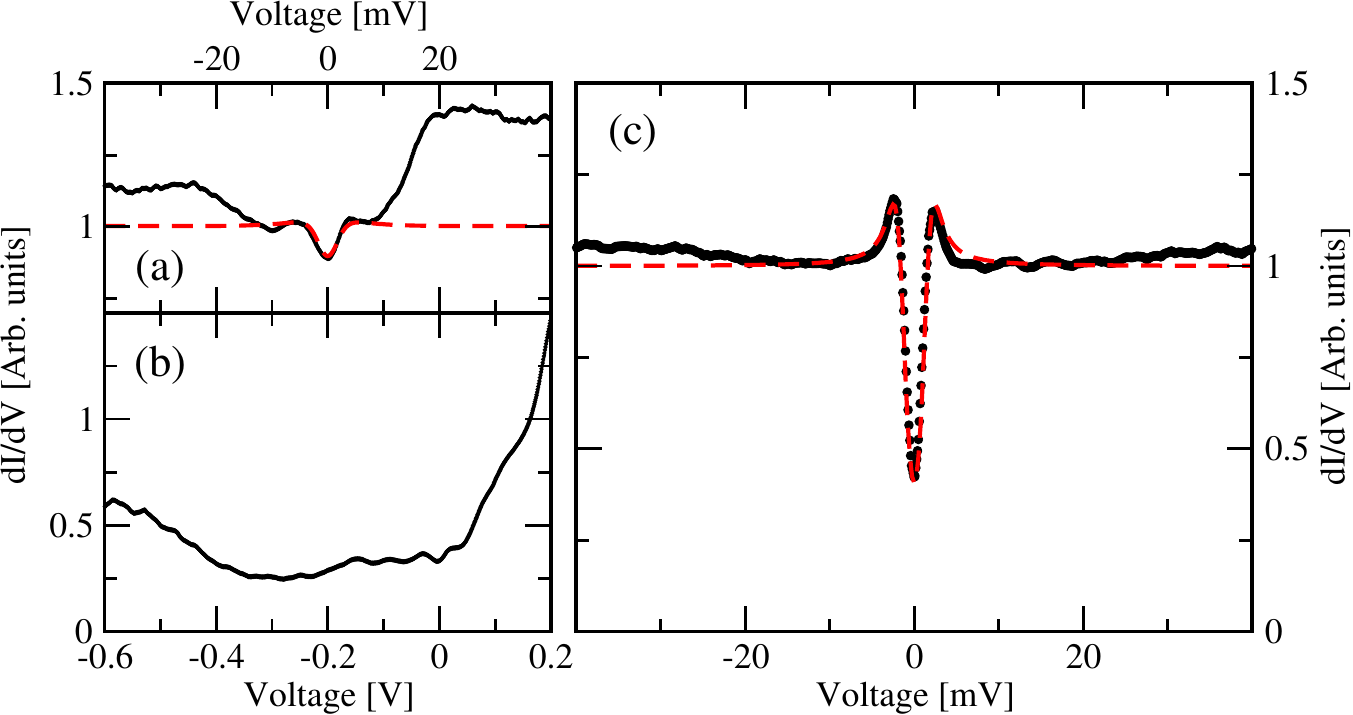}
\caption{[Color online] Measurements on samples of type A.
Panel (a):  $dI/dV$ curves taken around 0 V on 40 nm Nb on Bi$_2$Se$_3$.  Two gaps are apparent.  The larger gap is consistent with the aforementioned gap-like feature in figure 2(a).  The interior gap appears to be of superconducting origin and is fitted with a BCS $s$-wave superconducting energy gap fit.  The fitting parameters are:  $T=4.2$ K, $\Delta=1.50$ meV (3 mV peak to peak), mean free path $\ell=32$ nm.
Panel (b):  Wide range $dI/dV$ measurement done on 40 nm Nb on Bi$_2$Se$_3$.  The local minimum at approximately -300 mV (inset) is attributed to the Dirac point of the underlying Bi$_2$Se$_3$.  The gap-like feature at 0 V is too large to be superconductivity and likely arises from the band structure of electrons near the conduction band edge of the underlying Bi$_2$Se$_3$.
Panel (c):  The black curve is experimental data taken on Nb with Al$_2$O$_3$ sub-layer.  The dashed red curve is the same BCS fitting function used in figure 2(b), with the following parameters: $T=4.2$ K, $\Delta=1.50$ meV, $\ell=200$ nm.  The slight background slope points to a tunneling barrier on the scale of a few eV, consistent with the vacuum tunneling barrier formed by Nb (4 eV work function) and PtIr (5 eV work function).}
\label{ref:figure2}
\end{figure}

Samples of both types were studied.  Samples of type A have a pristine sub-layer of Bi$_2$Se$_3$ that was prepared in vacuum, resulting in a Dirac point around -300 mV, as expected.  The superconducting energy gap of the over-layer Nb is always centered on the Fermi level, 0 mV, well away from the Dirac point.  This asymmetry between the salient features, the Dirac point and the superconducting energy gap, lets us resolve both effects separately on the same measurement.  Scanning tunneling spectroscopy (STS) was performed first over a stripe of Nb on AlO$_x$. A superconducting energy gap with a peak-to-peak width of 3 mV was measured and fit using an $s$-wave BCS fitting function [see figure \ref{ref:figure2}(c)]. This measurement served as a quality check for our PtIr STM tip before moving on to a stripe of Nb with underlying Bi$_2$Se$_3$. At this location, both wide and narrow voltage ranges were explored. Figure \ref{ref:figure2}(b) shows a wide voltage range study of 40 nm of Nb on top of Bi$_2$Se$_3$. Comparison of these density of states (DOS) measurements to density of states measurements in the literature \cite{Hanaguri2010,Cheng2010} and done in our lab \cite{Urazhdin2004} for bare Bi$_2$Se$_3$ highlight two features.  First, a local minima [figure \ref{ref:figure2}(b)] is observed at -300 mV, which we interpret as the Dirac point of the underlying Bi$_2$Se$_3$. For positions without an underlying TI this minima is absent. A second gap-like feature at 0 V is consistent with the band structure of electrons near the conduction band edge seen in bare Bi$_2$Se$_3$. The gap at 0 V is not merely the superconducting energy gap, as this gap is approximately 45 mV in width.

Narrow voltage range measurements were then carried out around 0 mV near the Fermi level [figure \ref{ref:figure2}(a)]. Two gap-like features are evident. The smaller of the gaps appears to be of superconducting origin. To explore this, we fit the smaller gap with the same BCS fitting function used in figure \ref{ref:figure2}(c). The fitting parameters for figure \ref{ref:figure2}(a) employ more scattering, but the same temperature and energy gap as the fit for the superconductor with no underlying TI. The larger of the gaps is approximately 45 mV wide, and is consistent with the conduction band structure dip of the sub-layer Bi$_2$Se$_3$ observed in the long range DOS [figure \ref{ref:figure2}(b)]. Data on type A samples support the notion of the TPSS of the TI leaking through the superconductor, but cannot ultimately be fit with the theoretical model, as the model requires symmetry of the density of states about 0 V.  The superconducting gap is well resolved in figure \ref{ref:figure2}(a), in the absence of an underlying TI, therefore the additional gap-like feature in figure \ref{ref:figure2}(b) must be due to the underlying TI, and is likely caused by the gap edge of the conduction band of the TI.

\begin{figure}
\includegraphics[width=0.8\columnwidth]{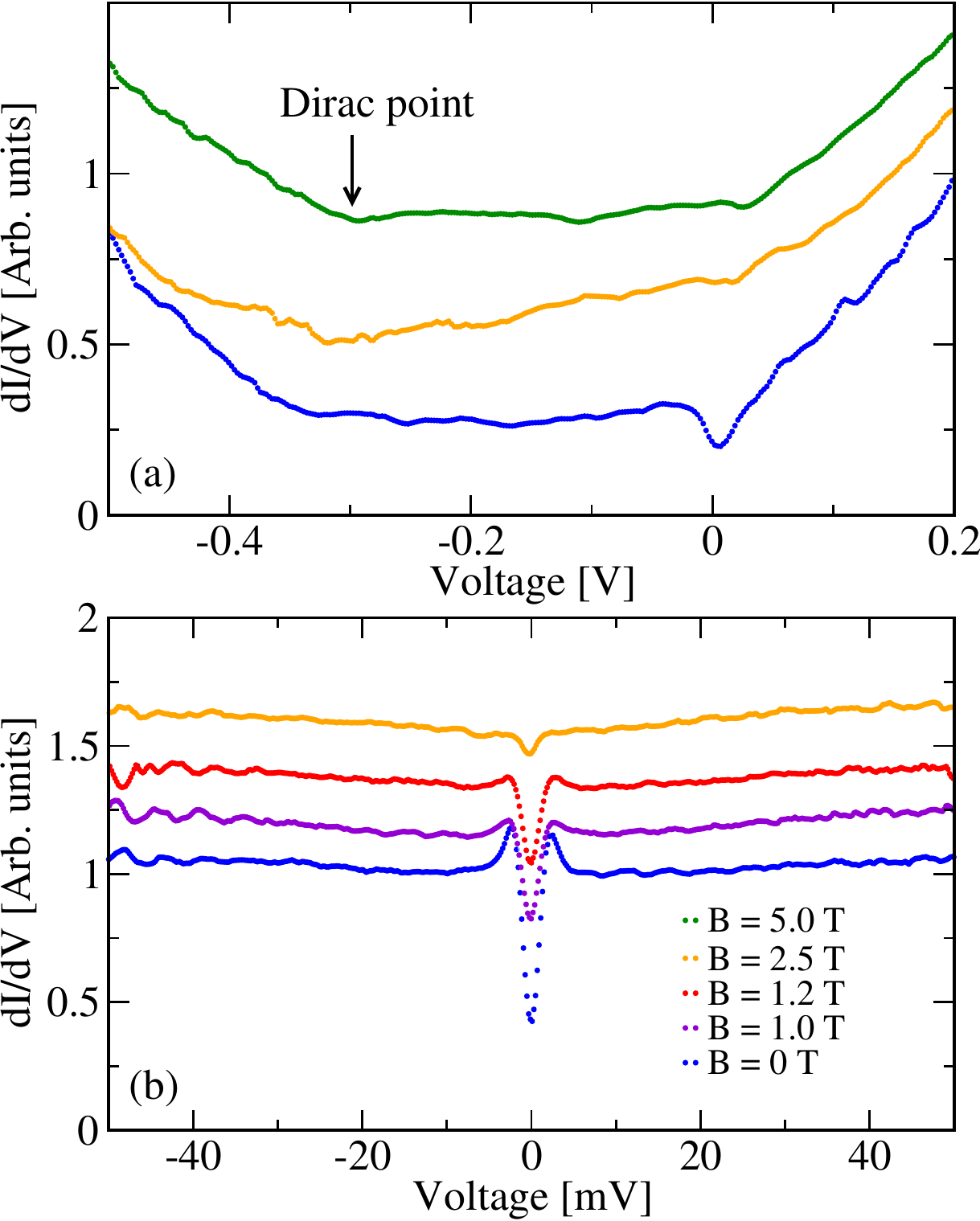}
\caption{[Color online] Measurements on samples of type A. Magnetic field spectroscopic measurements acquired on 60 nm Nb on (a) Bi$_2$Se$_3$, and (b) on Al$_2$O$_3$.  The coherence peaks and gap depth diminish as the field increases, with superconductivity ultimately suppressed near 2.5 T.  These critical field values are in agreement with transport measurements performed on similar Nb films. Wide voltage measurements in panel (a) show that although the extra gap like feature at zero bias is diminished with increasing magnetic field, it still persists, consistent with it having a non-superconducting origin from the Bi$_2$Se$_3$ substrate which is amplified by superconductivity. Panel (b) shows narrow bias data, where the superconducting gap can clearly be seen to be removed at a sufficiently high applied field of approximately 2.5 T.}
\label{ref:figure3}
\end{figure}

Figure \ref{ref:figure3} shows the evolution of superconductivity as a function of magnetic field for 60nm Nb on (a) Bi$_2$Se$_3$ and (b) Al$_2$O$_3$.  Consistent with transport measurements on similar Nb thin films, the superconducting energy gap is diminished with increasing field, implying a critical field around 2.5 T, see figure \ref{ref:figure3}(b).  Figure \ref{ref:figure3}(a) shows the evolution of the larger gap feature at 0 V.  With no magnetic field applied, the gap-like feature is pronounced.  Repeated measurements at 2.5 T and 5.0 T continue to show a smaller persistent zero energy gap, indicating the feature at 0 T was due not only to superconductivity, but also the underlying Bi$_2$Se$_3$ band structure.  In other words, superconductivity is not responsible for this gap-like feature, but rather enhances the underlying Bi$_2$Se$_3$ band structure which leaks to the surface.
 
 This comparison shows that the relative spectral weight of the Dirac cone and semiconductor gap edge features varies with Nb thickness. Additional measurements will be needed to characterize the length scale of the leakage of Dirac cone states. The TPSS can be expected to penetrate into the Nb over a length scale consistent with the Nb mean free path, hence a characteristic length scale of tens of nanometers is reasonable. From the fit of the BCS gap in Figure \ref{ref:figure2}(c) this is $\ell=200$nm, though this fitting procedure will also include other effects which broaden the density of states, and therefore overestimates the mean free path. We still see clear evidence of the dual proximity effect even for thicknesses of 60 nm. The effect will naturally gradually weaken as the thickness is increased. We note that one can not use the mean free path extracted from the fit in \ref{ref:figure2}(a) as an estimate of the Nb mean free path as it will also include effects from the hybridization of the BCS and TPSS states, see section \ref{sec_eff_bcs}.

\subsection{Samples of Type B}

\begin{figure}
\includegraphics[width=0.99\columnwidth]{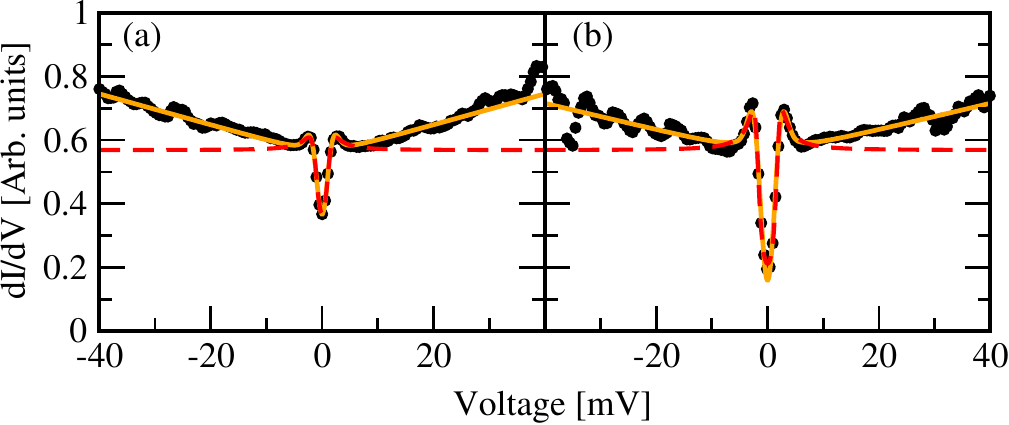}
\caption{[Color online] Measurements on samples of type B. The measured differential conductance, black circles, for three examples of Nb on top of the TI. Fits to the density of states for the dispersion Eq.~\eqref{spectrum} are shown in solid orange, and fits to the BCS theory are shown in dashed red. As can be seen only the full model correctly captures the structure seen in the experiment. Fitting parameters for the full theory are: (a) $\Delta=0.867$ meV, $v_F=395$ ms$^{-1}$, $G=1.45$ meV, and $\gamma=2.41$ meV; and (b) $\Delta=1.58$ meV, $v_F=427$ ms$^{-1}$, $G=1.47$ meV, and $\gamma=3.69$ meV.}
\label{ref:figure4}
\end{figure}

Samples of type B are symmetric with respect to the Dirac cone and superconducting energy gap about zero, and serve as ideal candidates for testing the theoretical model. We compare the differential conductance measured experimentally on three areas of 30 nm Nb on Bi$_2$Se$_3$ to the theory. At low temperature the differential conductance measured in the experiments is then $dI/dV\approx\nu(eV)$. In figure \ref{ref:figure4} we show several fits using $\mu_{\rm TPSS}=0$, and $v_F$, $\Delta$, $\gamma$, and $\Gamma$ as fitting parameters. The overall magnitude of the density of states is also a fitting parameter. We find very good fits to both the gap structure and the Dirac cones. From these fits one has $O(\Delta^{\rm eff})=O(\gamma)=O(\Delta)$, so the induced superconductivity for the TPSS is still reasonable. However although the magnitude of $\gamma$ is found to be of the order of meV, the fits are not sufficient to pinpoint it precisely, and a certain range of possible values could still be consistent with the data. Also shown are fits to the standard BCS theory, which naturally can not fit the Dirac cone like features which we see. We also considered fits to equation \eqref{dosfk}, not shown, but they do not capture the gap feature. We have observed similar curves on a variety of sample thicknesses of Bi$_2$Se$_3$, and the effect is visible even up to 60 nm layers. As the sample thickness is increased we expect the effective Fermi velocity in our model to decline, however sample to sample variations made it impossible to discern a clear pattern in this.

However we were not able to discern a length scale to the dual proximity effect.

The gap $\Delta$ used in the fits is slightly smaller than the known value for bulk Nb, which is $3.05$meV. This is to be expected as the thin film can only just be considered a bulk sample, and additionally the presence of the TPSS will have an effect in reducing the strength of the pairing. The Fermi velocity of the TPSS is also reduced from its value on the Bi$_2$Se$_3$ surface, in this case drastically. The velocity on the clean surface is $v_F^{\rm cl}=5\cdot10^5$ms${}^{-1}$. According to the fits in figure \ref{ref:figure4} it is reduced by three orders of magnitude, caused by the states widening throughout the Nb layer. From a simple hydrodynamic consideration we would expect the velocity for the TPSS in the Nb to be approximately $v_F=v_F^{\rm cl}\xi/(d+\xi)$, where $d$ is the thickness of the Nb and $\xi\approx 5$nm is the confinement of the TPSS in Bi$_2$Se$_3$. This gives $v_F\approx1200$ms$^{-1}$, which is three times larger than the measured values. This discrepancy we put down to hybridization effects beyond our phenomenological model.

A smaller Fermi velocity for the TPSS in turn means that their coherence length will be smaller, and hence any MBS will be confined very sharply near the vortex cores. Curiously, strong localization of MBS was recently found in a different system of ferromagnetic atomic chains deposited on a surface of a superconductor~\cite{Nadj-Perge2014}. It was understood that short coherence length of MBS results from the strong Fermi velocity renormalization caused by a quasiparticle weight shift of the electrons' spectral weight from the adatom-wire into the SC via local hybridization mechanism. We reveal that a similar effect of strong velocity renormalization takes place in our proximity system. 

\section{Summary}

We have investigated the density of states on the surface of superconducting Nb deposited on a large Bi$_2$Se$_3$ substrate, finding some striking and unforeseen behavior. A Dirac cone is quite clearly visible, despite not being native to the Nb. This is due to the TPSS of the underlying Bi$_2$Se$_3$ leaking into the Nb, and hybridizing with its states. Our theoretical model fits very well to the experimental data, strengthening this proposed explanation. This is also in agreement with the results of Ref.~\onlinecite{Trang2020} where Pb films are grown on TlBiSe$_2$. While such a setup does still contain effective $p$-wave like pairing, resulting in a topological superconductor which could host MBS there are several caveats to the Fu-Kane model \cite{Fu2008} which arise. As the TPSS leak into the superconductor, a thick superconducting layer would destroy the 2D Dirac cone. This can be circumvented by either using thin layers, which are nonetheless thick enough to be bulk superconductors, or by inserting an insulating layer between the TI and the superconductor. This second scenario would however reduce the size of the proximity effect for the TPSS in the TI surface layers. We also note that the effective Fermi velocity for the TPSS is reduced by several orders of magnitude due to spreading into the superconductor, which changes the parameter space in which MBS could be found.

\phantom{.}

\emph{Comment: After our paper appeared on the arXiv we became aware of a similar study considering a model of a coupled superconductor thin film and topological insulator~\cite{Hugdal2019}.}

\acknowledgments

This work is supported by the U.S.~Department of Energy, Office of Science, Basic Energy Sciences, under Award DE-SC0017888. Additional support for work at the University of Wisconsin-Madison was provided by NSF CAREER Grant DMR-1653661, NSF EAGER Grant DMR-1743986, and by the Wisconsin Alumni Research Foundation.  At the University of Illinois at Urbana-Champaign, C.Z. was supported for device fabrication by NSF DMR-1610114, and we acknowledge the use of the central facilities of the Frederick Seitz Materials Research Laboratory.  Work at Argonne National Laboratory was supported by the U.S. Department of Energy, Office of Science, Basic Energy Sciences under Contract No.~DEAC02-06CH11357. In M.Curie-Sk{\l}odowska University N.S.~was supported by the National Science Centre (NCN, Poland) under Grant No.~UMO-2018/29/B/ST3/01892.

\end{document}